\documentclass[%
 reprint,
 superscriptaddress,
 amsmath,amssymb,
 aps,
 longbibliography,
]{revtex4-1}

\usepackage[dvipdfmx]{graphicx}
\usepackage{dcolumn}
\usepackage{bm}
\usepackage{color}
\usepackage{ulem}

\begin{document}


\title{
Anomalous electrical magnetochiral effect by chiral spin fluctuation
}

\author{Hiroaki Ishizuka}\email{ishizuka@appi.t.u-tokyo.ac.jp}
\affiliation{
Department of Applied Physics, The University of Tokyo, Bunkyo, Tokyo, 113-8656, JAPAN 
}

\author{Naoto Nagaosa}
\affiliation{
Department of Applied Physics, The University of Tokyo, Bunkyo, Tokyo, 113-8656, JAPAN 
}
\affiliation{
RIKEN Center for Emergent Matter Sciences (CEMS), Wako, Saitama, 351-0198, JAPAN
}

\begin{abstract}
The non-collinear spin configurations cause many nontrivial phenomena related to the Berry phase. They are described by the vector spin chirality $\bm\chi_{ij} ={\bf S}_i\times{\bf S}_j$ or scalar spin chirality $\chi_{ijk}=({\bf S}_i \times {\bf S}_j) \cdot {\bf S}_k$, which are related to the spin current and effective magnetic field, respectively. The scalar spin chirality leads to the topological Hall effect in metals~\cite{Ye1999,Ohgushi2000}, while the vector spin chirality to the ferroelectricity of spin origin, i.e., multiferroics in insulators~\cite{Katsura2005}. However, the role of the vector spin chirality in conducting systems has not yet elucidated. Here we show theoretically that the spin fluctuation with vector spin chirality in chiral magnets scatters electrons asymmetrically, resulting in a nonreciprocal transport phenomena, i.e., electrical magnetochiral effect (eMChE)~\cite{Rikken2001}. This asymmetric scattering appears in the leading-order scattering term, implying a large nonreciprocity in the charge and spin currents. We find that the temperature and magnetic field dependence of the eMChE reproduces that observed in MnSi~\cite{Yokouchi2017}. Our results reveal the microscopic mechanism of eMChE and its potential in producing a large nonreciprocal response.
\end{abstract}

\date{\today}

\maketitle

Vector $\bm\chi_{ij}$ and scalar $\chi_{ijk}$ spin chiralities are central concepts in the physics of non-collinear spin structures. Since the spin operator is odd in ${\cal T}$,  $\chi_{ij}$ is even while $\chi_{ijk}$ is odd. Therefore, $\chi_{ijk}$ is related to the magneto-transport; topological Hall effect associated with $\chi_{ijk}$, both intrinsic~\cite{Ye1999,Ohgushi2000} and extrinsic~\cite{Tatara2002,Ishizuka2018} mechanisms, are studied~\cite{Nagaosa2010}. On the other hand, the inversion symmetry operation $P$ about the center of the bond connecting $i$ and $j$ reverses the sign of $\bm\chi_{ij}$. The symmetry property implies  $\bm\chi_{ij}$ is related to the electric polarization of spin origin in insulators~\cite{Katsura2005}. In conducting systems, on the other hand, the broken $P$ is subtle since the electric field in the metal is prohibited. However, there are several interesting nonreciprocal transport phenomena in noncentrosymmetric crystals~\cite{Rikken2001,Rikken2005,Tokura2018}.  

The reciprocal theorem by Onsager provides a basis to discuss the nonreciprocal linear responses~\cite{Onsager1931,Kubo1957}. This theorem originates from the time-reversal symmetry ${\cal T}$ of the microscopic dynamics, which is different from the macroscopic irreversibility in the macroscopic scale. In transport theory, the Hermite symmetry also gives the reciprocal relation~\cite{Fisher1981}, in addition to the space group symmetry of the crystal. Therefore, the breaking of $P$ alone does not necessarily lead to nonreciprocal responses. The nonreciprocity becomes even more subtle and rich for the nonlinear responses~\cite{Rikken2001,Rikken2005}. The nonreciprocal dc transport in solids manifests in the $I^2$ term of the $I$-$V$ curve, where $I$ is the injected electric current and $V$ is the voltage drop~\cite{Rikken2001,Rikken2005}. For example, the $I$-$V$ curve of eMChE follows~\cite{Rikken2001}
\begin{align}
V=R_0(1+\gamma(B) IB)I,\label{eq:Rikken}
\end{align}
which means that the external magnetic field is needed to break ${\cal T}$ for this effect.  Recent experiments found the nonreciprocal response in various nonmagnetic materials such as Bi helix~\cite{Rikken2001}, semiconductors subject to gate potential~\cite{Rikken2005}, molecular conductors~\cite{Pop2014}, polar semiconductor~\cite{Ideue2017}, and superconductor~\cite{Wakatsuki2017}. The nonreciprocal response also appears in magnetic materials, such as eMChE in metal/ferromagnet bilayer~\cite{Avci2015}, magnetic topological insulator~\cite{Yasuda2016}, and chiral magnets~\cite{Yokouchi2017,Aoki2019}. In the magnetic systems, the magnetic ordering and fluctuation seem to play a crucial role in sharp contrast to the band structure effects dominating in the nonmagnetic systems. Among them, a recent paper reported a detailed experiment on the temperature and magnetic field dependence of eMChE in MnSi~\cite{Yokouchi2017}, providing a useful set of information for theoretical studies.  MnSi is a chiral magnet with a helical magnetic order in the zero field~\cite{Ishikawa1976,Bak1980}. This material and its sister compounds are known for the magnetic-skyrmion crystal phase~\cite{Roessler2006,Muhlbauer2009,Yu2010}. A recent experiment finds that MnSi also shows nonreciprocal response similar to the eMChE but with a non-monotonic magnetic field dependence~\cite{Yokouchi2017}; similar behavior also appears in CrNb$_3$S$_6$~\cite{Aoki2019}. These papers report a non-monotonic temperature dependence of the eMChE, which shows a maximum at around the magnetic transition temperature. The result implies the importance of magnetic fluctuation. However, the microscopic mechanism on how the magnetic fluctuation produces nonreciprocity remains elusive.

\begin{figure}
  \includegraphics[width=\linewidth]{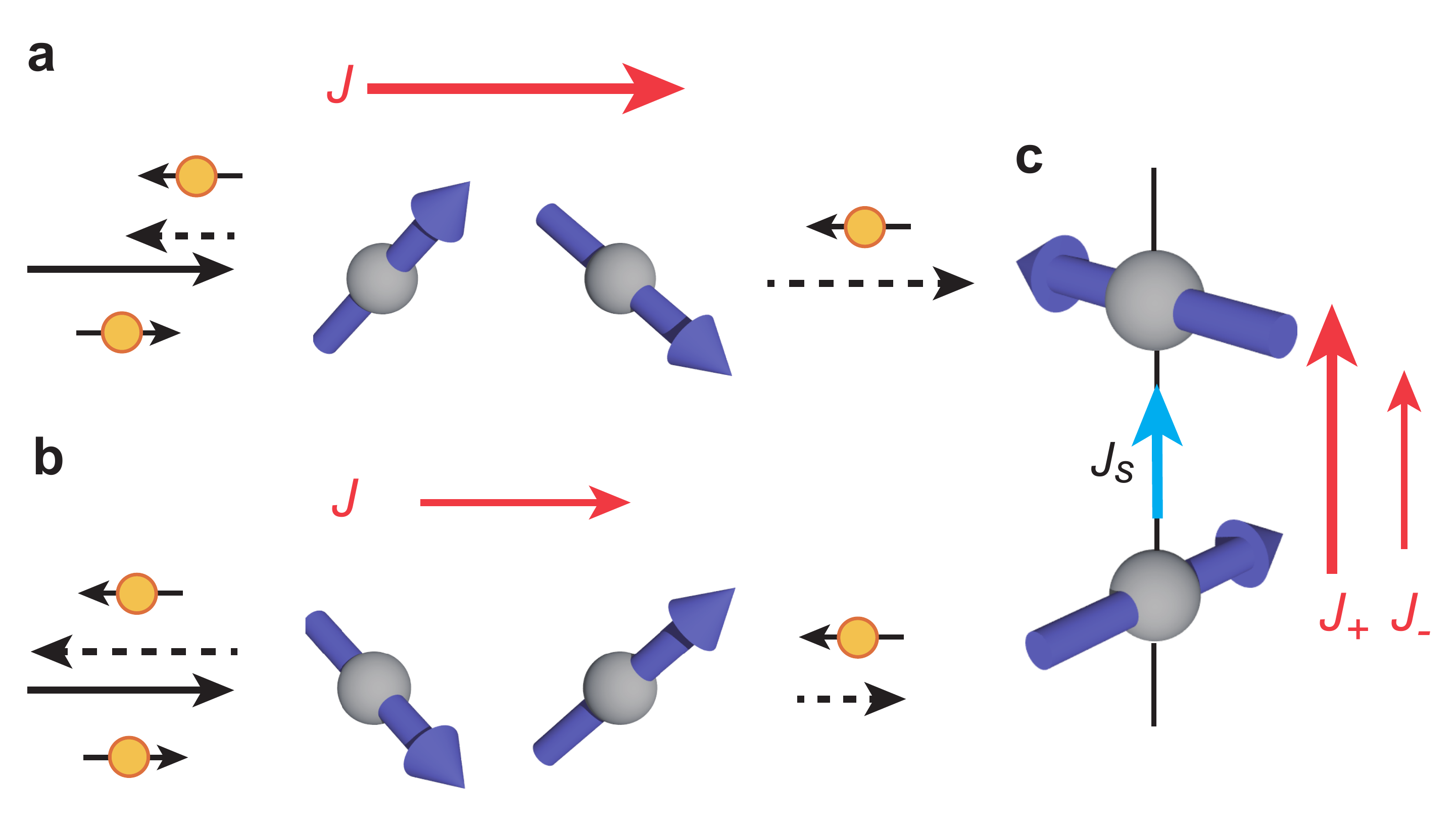}
  \caption{{\bf Nonreciprocal transport by magnetic scattering.} (a,b) Schematic figure of magnetic scattering by a two-spin cluster with finite vector spin chirality. The backward scattering by the two-spin spin cluster scatters incoming electron with ``up'' spin (the electron at the light bottom of the figure) depends on the vector spin chirality; less electrons are scattered backward in (a) compared to (b). The weaker backward scattering in (a) results in a larger current compared to (b). (c) Nonreciprocal spin current for $S^z$ in the paramagnetic case. The spin current of electrons flow along the direction of the ``supercurrent'' of magnetic moments because of the difference between the current for up-spin electrons and down-spin ones.}
  \label{fig:intro:schematic}
\end{figure}

In this work, we theoretically study the nonreciprocal transport phenomena of electrons focusing on an asymmetric scattering by magnetic fluctuations. We find the magnetic fluctuation in chiral magnets causes asymmetric scattering in the leading order of the scattering. The asymmetry produces a nonreciprocal response of electric current in nonmagnetic systems~\cite{Isobe2018}. In contrast to Ref.~\cite{Isobe2018}, the magnetic fluctuation produces a larger nonreciprocal current because the asymmetric scattering appears in the leading order. Using the semiclassical Boltzmann theory, we show the magnitude of the nonreciprocal current is consistent with that in the experiment. Moreover, the temperature and magnetic field dependences reproduce the experiment. The consistency between the experiment and our theoretical results provides strong evidence of the ``extrinsic'' mechanism for nonreciprocal electric current. 

\section*{Results}

\subsection*{Asymmetric scattering rate}

To study how the electron scattering produce nonreciprocal response, we here consider a model with itinerant electrons and localized spins coupled by exchange interaction. The Hamiltonian is
\begin{align}
H=&\sum_{\bf k\sigma}\varepsilon_{{\bf k}\sigma} c_{\bf k\sigma}^\dagger c_{\bf k\sigma}+\frac{J}N\sum_{\substack{i, {\bf k}\\\alpha,\beta}} \gamma_{i\bf k}{\bf S}_i\cdot c_{{\bf k}\alpha}^\dagger \bm\sigma_{\alpha\beta} c_{{\bf k}\beta},\label{eq:result:Hamil}
\end{align}
where $c_{\bf k\sigma}$ ($c_{\bf k\sigma}^\dagger$) are respectively the annihilation (creation) operator of itinerant and localized electrons, $\bm \sigma\equiv(\sigma^x,\sigma^y,\sigma^z)$ is the vector of Pauli matrices $\sigma^a$ ($a=x,y,z$), $\varepsilon_{{\bf k}\sigma}=k^2/(2m)-\sigma M-\mu$ is the eigenenergy of itinerant electrons with momentum $\bf k$ and spin $\sigma=\pm1$ ($+1$ for up spin and $-1$ for down spin), $k\equiv|\bf k|$, $\gamma_{i\bm k}\equiv e^{{\rm i}\bm k\cdot\bm r_i}$, $J$ is the Kondo coupling between the localized spins and the itinerant electrons, and ${\bf S}_i$ is the localized moment at ${\bf r}_i$. Here, we assumed the magnetization is along the $z$ axis. This model is a classical spin Kondo lattice model if the localized spins exists on every site, and is a Kondo impurity model if the spins exist only on a few sites $N_s\ll N$.

We calculate the scattering rate of electrons by the localized spins using Born approximation. In the first Born approximation, the scattering rate $W_{\bf k\sigma,\bf k'\sigma'}$ of electrons from the ${\bf k}\sigma$ state to the ${\bf k}'\sigma'$ state reads:
\begin{align}
W_{{\bf k}\sigma,{\bf k}'\sigma'}
=&\frac{2\pi J^2}{N^2}\sum_{\substack{i,j\\a,b}}S_i^aS_j^b\sigma^a_{\sigma\sigma'}\sigma^b_{\sigma'\sigma} e^{{\rm i}({\bf k}'-{\bf k})\cdot({\bf r}_i-{\bf r}_j)}\delta(\varepsilon_{{\bf k}\sigma}-\varepsilon_{{\bf k}'\sigma'}).\label{eq:result:Wkk}
\end{align}
Here we assume that the spin fluctuation is classical and static, which is justified when the temperature is much higher than the typical energy of spin fluctuation. The experimental situation in MnSi discussed below satisfies this condition. A recent work point outs that the asymmetry in the scattering rate $W_{\bf k,\bf k'}\ne W_{-\bf k,-\bf k'}$ produces the nonreciprocity in the electron transport~\cite{Isobe2018} (Ref.~\cite{Isobe2018} considered spinless fermions.). Therefore, we focus on a similar asymmetry in $W_{\bf k\sigma,\bf k'\sigma'}$. The asymmetric part of the scattering rate ($W^-_{{\bf k}\sigma,{\bf k}'\sigma'}\equiv=(W_{{\bf k}\sigma,{\bf k}'\sigma}-W_{-{\bf k}\sigma,-{\bf k}'\sigma})/2$) reads
\begin{align}
W^-_{{\bf k}\sigma,{\bf k}'\sigma'}=&\frac{2\pi J^2}{N^2}\sigma\delta_{\sigma,\bar\sigma'}\sum_{i,j}^{N_s}\sin\left(({\bf k}-{\bf k}')\cdot{\bf r}_{ij}\right)\nonumber\\
&\qquad\qquad\qquad\times({\bf S}_i\times{\bf S}_j)_z\delta(\varepsilon_{\bf k\sigma}-\varepsilon_{\bf k'\sigma'}).\label{eq:result:Wakk}
\end{align}
Here, ${\bf r}_{ij}={\bf r}_i-{\bf r}_j$ and $\bar\sigma=-\sigma$; we assumed $\varepsilon_{\bf k\sigma}=\varepsilon_{-\bf k\sigma}$. This asymmetric scattering vanishes when $N_s=1$; the sine function is always zero because ${\bf S}_{1}\times{\bf S}_{1}=\bf0$.  Therefore, multiple spin scattering is necessary for the non-zero asymmetric scattering.

In the two spin case, the scattering rate reads
\begin{align}
W^-_{\vec k\sigma,\vec k'\sigma'}=\frac{4\pi J^2}{N^2}\sigma\delta_{\sigma,\bar\sigma'}&\sin\left(({\bf k}-{\bf k}')\cdot{\bf r_{12}}\right)\nonumber\\
&\times\left({\bf S}_1\times{\bf S}_2\right)_z\delta(\varepsilon_{\bf k\sigma}-\varepsilon_{\bf k'\sigma'}).\label{eq:result:W2spin}
\end{align}
Hence, the asymmetry appears when a non-zero vector spin chirality exists, i.e., when the two spins are non-collinear. A previous work on multiferroics in insulators point outs the relation of the local spin current $\propto {\bf S}_i\times{\bf S}_j$ and electric polarization~\cite{Katsura2005}. From a similar viewpoint, our result shows the local spin current scatters electrons asymmetrically depending on the spins [Fig.~\ref{fig:intro:schematic}(c)]. In addition, the result implies a finite magnetization is necessary for the nonreciprocity because the asymmetric scattering rate in Eq.~\eqref{eq:result:W2spin} has the opposite signs for $W^{-}_{{\bf k}\uparrow,{\bf k}'\downarrow}$ and $W^{-}_{{\bf k}\downarrow,{\bf k}'\uparrow}$. Therefore, the asymmetry cancels when the itinerant electrons are paramagnetic ($M=0$). In short, the above result implies nonreciprocity in the conductivity appears in a magnet when both the vector spin chirality and magetization are nonzero.

We note that $W^-_{{\bf k}\sigma,{\bf k}'\sigma'}$ appears in the first Born approximation. This feature is in contrast to Ref.~\cite{Isobe2018} where $W^-_{{\bf k}\sigma,{\bf k}'\sigma'}$ appears from the second Born terms, i.e., higher-order in the perturbation. For the non-magnetic scatterers in time-reversal symmetric system, $W^-_{{\bf k}\sigma,{\bf k}'\sigma'}$ is related to the skew scattering $W^s_{{\bf k},{\bf k}'}=(W_{{\bf k},{\bf k}'}-W_{{\bf k}',{\bf k}})/2$ by ${\cal T}$~\cite{Isobe2018}. The skew scattering is prohibited in the first Born approximation because of the Hermiticity of the impurity potential. Therefore, the second-order term is the leading order. In contrast, the magnetic scattering considered here breaks ${\cal T}$. This difference of the symmetry allows non-zero $W^-_{{\bf k}\sigma,{\bf k}'\sigma'}$ in the leading-order first Born approximation. This result also implies that the magnetic scattering produces a larger nonreciprocal response. 

\subsection*{Boltzmann theory for nonreciprocal currents}

To study how $W^-_{{\bf k}\sigma,{\bf k}'\sigma'}$ contributes to the eMChE, we calculate the conductivity using the semiclassical Boltzmann theory.  Using the relaxation time approximation, the Boltzmann equation reads
\begin{align}
e{\bf E}\cdot{\bf \nabla}_{k}f_{{\bf k}\sigma}
=&-\frac{f_{{\bf k}\sigma}-f^0_{{\bf k}\sigma}}\tau+\sum_{{\bf k}',\sigma'}W_{{\bf k}\sigma,{\bf k}'\sigma'}^{-}(f_{{\bf k}'\sigma'}-f_{{\bf k}\sigma}).\label{eq:result:boltzmann}
\end{align}
Here, $e<0$ is the elementary charge, ${\bf E}=(E_x,E_y,E_z)$ is the applied electric field, and $f_{{\bf k}\sigma}$ is the electron density for the electrons with momentum ${\bf k}$ and spin $\sigma$. For simplicity, we focus on the case ${\bf E}=(0,0,E)$. In addition, we assume
\begin{align}
W_{{\bf k}\sigma,{\bf k}'\sigma'}^{-}=
\left\{\begin{array}{ll}
0 & \qquad\text{(if $\sigma=\sigma'$)}\\
2\pi \sigma c (k_z-k'_z)\delta(\varepsilon_{\bf k\sigma}-\varepsilon_{\bf k'\sigma'}) & \qquad\text{(if $\sigma\ne\sigma'$)}
\end{array}\right.,\label{eq:result:Wasym}
\end{align}
where $c=\frac{J^2}{N}\chi_v$ is a real constant and $\chi_v\equiv\langle({\bf S}_i\times {\bf S}_j)_z\rangle$ is the thermal average of the $z$ component of the vector spin chirality between the nearest-neighbor spins along the $z$ axis. This asymmetric scattering term corresponds to the thermal average of the $k\ll1$ case of the two-spin impurity cluster in Eq.~\eqref{eq:result:W2spin}.

We solve the Boltzmann equation in Eq.~\eqref{eq:result:boltzmann} with the scattering rate in Eq.~\eqref{eq:result:Wasym} by expanding $f_{{\bf k}\sigma}$ up to the second order in $\bf E$ and linear order in $c$~\cite{Isobe2018}. Within this approximation, the nonreciprocal current reads
\begin{align}
J_z^{(2)}=-\frac{144\pi}5\frac{\tau m}{e\mu^2}cM\sigma_0^2E^2,\label{eq:result:S2lowM}
\end{align}
where $2\sigma_0=\frac{4e^2\tau\mu}{3mn}$ is the linear conductivity of electrons at $M=0$ and $c=0$. Here, $n$ is the density of state at the Fermi level and we assumed $\mu\gg M$. Hence, the scattering by the two spins produce non-reciprocal current proportional to $c$ and magnetic polarization of the itinerant electrons $M$.

We also note that the two spin scattering produces the spin current. Using the same formalism, we find the spin current for $S^z$ reads
\begin{align}
J^z_z=-\frac{54\pi\hbar}5\frac{\tau m}{e^2\mu} c\, \sigma_0^2 E^2.
\end{align}
Unlike the charge current, the spin current appears without the spin polarization. Therefore, a paramagnet with the chiral spin fluctuation produces a finite spin current by simply flowing electric current. 

\subsection*{Nonreciprocal charge current in chiral magnets}

\begin{figure}
  \includegraphics[width=\linewidth]{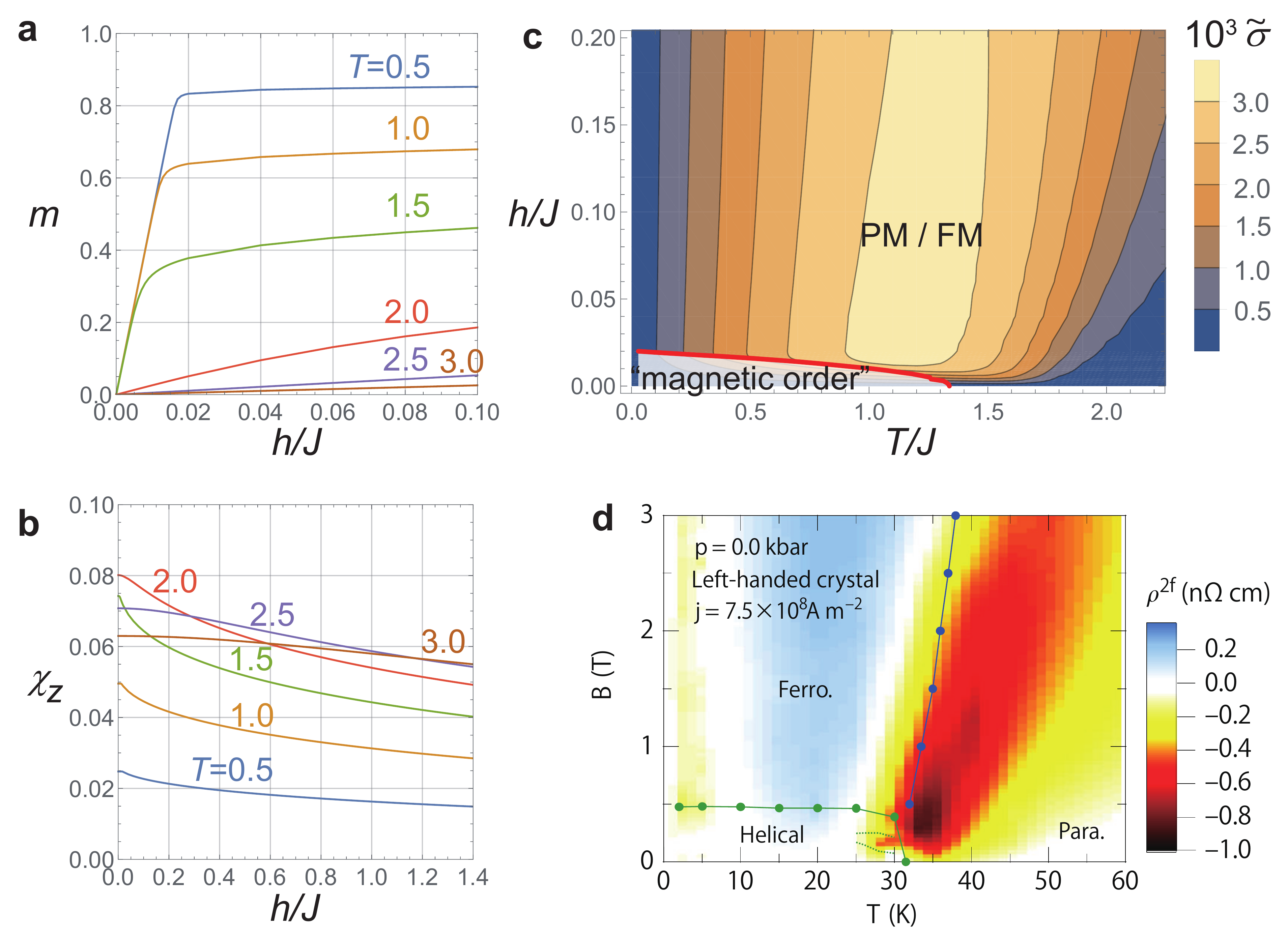}
  \caption{{\bf Magnetic and transport properties of a three-dimensional chiral magnet.} The magnetic-field dependence of (a) magnetization $M$ and (b) vector spin chirality $\chi_z$ for different temperature $T$. (c) is the contour plot of $\tilde\sigma=M\chi_z$. The results are for $D/J=0.2$. The red line is the phase boundary between the ordered and paramagnetic (PM/FM) phases, which is determined by $\lambda+D^2/(4J)>-10^{-4}$. See Method section for details. (d) The contour plot of second harmonic resistivity $\rho^{2f} (\propto \sigma^{(2)})$. Reproduced from Ref.~\cite{Yokouchi2017}.}
  \label{fig:result:MnSi}
\end{figure}

In the above mechanism, the nonreciprocal current depends on temperature and magnetic field via that of the magnetization and vector spin chirality. To investigate the dependence of nonlinear conductance, we here consider a classical ferromagnetic Heisenberg model on a cubic lattice with Dzyaloshinskii-Moriya interaction~\cite{Bak1980,Yu2010},
\begin{align}
H_{\rm cm}=-J\sum_{\langle i,j\rangle}{\bf S}_i\cdot{\bf S}_j-\frac{D}2\sum_{\langle i,j\rangle}{\bf r}_{ji}\cdot{\bf S}_i\times{\bf S}_j-h\sum_i S_i^z.
\end{align}
Here, the sum is over the nearest-neighbor bonds. Figure~\ref{fig:result:MnSi} shows the magnetic and transport properties of the above model; all results are obtained using Onsager's reaction field theory. Figure~\ref{fig:result:MnSi}(a) is the plot of the magnetization to the magnetic field. The result shows a ferromagnetic magnetization curve below $T\lesssim 2J$ due to the ferromagnetic $J$, which decreases monotonically with increasing temperature. In contrast, the vector spin chirality $\chi_z$ shows a non-monotonic temperature dependence. Figure~\ref{fig:result:MnSi}(b) shows the magnetic field dependence of $|\chi_z|$ for different $T$. When $T/J\lesssim 1$, the field-induced magnetization suppresses $|\chi_z|\to0$ as $T\to0$. With increasing temperature, the thermal fluctuation increase $|\chi_z|$ by suppressing the magnetization. The maximum is around $T/J\sim2-3$ depending on $h$; the maximum tends to move to a higher $T$ as $h$ increases. Further increase of the temperature reduces the spin chirality because the thermal fluctuation dominates over the exchange interactions between the spins. 

Equation~\eqref{eq:result:S2lowM} shows the nonreciprocal conductivity $\sigma^{(2)}$ is proportional to $\tilde\sigma^{(2)}\equiv M\chi_z$. Figure~\ref{fig:result:MnSi}(c) shows the contour plot of $\tilde\sigma^{(2)}$ in the $T-h$ plane. In the low temperature region, the result shows a small $\tilde\sigma^{(2)}$ owing to the suppression of the vector chirality. With increasing the temperature, $\tilde\sigma^{(2)}$ increases due to the increase of $\chi_z$ with a maximum around $T/J\sim 1.5-2$; $\tilde\sigma^{(2)}$ then decreases because both $M$ and $\chi_z$ is suppressed by the thermal fluctuation when $T/J\gg1$. Figure~\ref{fig:result:MnSi}(c) also shows the increase of the maximum with increasing the magnetic field. This is related to the increase of the maximum of $\chi_z$ discussed above. These trends are qualitatively consistent with the experiment in MnSi as shown in Fig.~\ref{fig:result:MnSi}(d)~\cite{Yokouchi2017}.

\section*{Discussions}

To summarize, we studied the nonreciprocity of electric current produced by the fluctuation of localized spins. We find that the scattering process involving two spins cause an asymmetric scattering, which is proportional to the vector spin chirality. This effect appears at the leading order in the impurity scattering, i.e., within the first Born approximation. Therefore, we expect a large asymmetry in the scattering rate. Using the semiclassical Boltzmann theory, we find that this asymmetry produces nonreciprocal transport of electrons; $\sigma^{(2)}$ is proportional to the vector spin chirality and spin polarization of itinerant electrons. We also find the chiral spin fluctuation produces nonreciprocal spin current. As a consequence, $\sigma^{(2)}$ shows a non-monotonic temperature with a maximum around $T/J\sim1$. This trend is consistent with the recent experiments in MnSi~\cite{Yokouchi2017} and CrNb$_3$S$_6$~\cite{Aoki2019}. In particular, the overall behavior of $\sigma^{(2)}$ well accounts for the eMChE in MnSi~\cite{Yokouchi2017}.

The magnitude of the eMChE by the magnetic scattering is also consistent with the experiment in MnSi~\cite{Yokouchi2017}. A recent experiment of MnSi finds the ratio of linear and nonreciprocal resistivities $\gamma(B)IB\sim 10^{-4}-10^{-5}$ with $I=10^9$ Am$^{-2}$. When $\gamma(B)IB\ll1$, the ratio reads $\gamma IB\sim -\sigma^{(2)}I/\sigma^2$. We estimate $\sigma^{(2)}$ using Eq.~\eqref{eq:result:S2lowM} assuming $J=10$ meV, $D=1$ meV, $a_0=4$ \AA, $m=9.109\times10^{-31}$ kg, $\rho=1/(2\mu_Fa_0^3)\sim10^{39}$ J$^{-1}$cm$^{-3}$, and $\tau=10^{-13}$ s. We use $\mu_F=0.5$ eV and $M(=g\mu_BH)=100$ meV because the bandwidth is $\sim1$ eV~\cite{Jeong2004} and the spin polarization is in the order of 10\%~\cite{Lee2007}. Using these values, we find $\sigma^{(2)}I/\sigma^2\sim2\times10^{-5}$. Therefore, the result is roughly comparable with that observed in MnSi.

\appendix
\section*{Method}

\subsection*{Boltzmann theory}

We used the semiclassical Boltzmann theory to calculate the nonreciprocal current. Assuming the steady state, the semiclassical Boltzmann equation reads
\begin{align*}
e{\bf E}\cdot{\bf \nabla}_{k}f_{{\bf k}\sigma}
=&\sum_{{\bf k}',\sigma'}\left(W_{{\bf k}\sigma,{\bf k}'\sigma'}f_{{\bf k}'\sigma'}-W_{{\bf k}'\sigma',{\bf k}\sigma}f_{{\bf k}\sigma}\right).
\end{align*}
Here, $e<0$ is the elementary charge, ${\bf E}=(E_x,E_y,E_z)$ is the applied electric field, and $f_{{\bf k}\sigma}$ is the electron density for the electrons with momentum ${\bf k}$ and spin $\sigma$. For simplicity, we focus on the case ${\bf E}=(0,0,E)$. The first (second) term in the right-hand side of the equation represents the scattering of electrons from ${\bf k}'\sigma'$ (${\bf k}\sigma$) to ${\bf k}\sigma$ (${\bf k}'\sigma'$).  We approximate the symmetric part of the scattering rate by a relaxation time $\tau$. A similar approximation were used elsewhere to study transport phenomena related to a specific scattering term~\cite{Ishizuka2018,Ishizuka2017,Ishizuka2018b,Isobe2018}. Within this approximation, the Boltzmann equation reads
\begin{align*}
e{\bf E}\cdot{\bf \nabla}_{k}f_{{\bf k}\sigma}
=&-\frac{f_{{\bf k}\sigma}-f^0_{{\bf k}\sigma}}\tau+\sum_{{\bf k}',\sigma'}W_{{\bf k}\sigma,{\bf k}'\sigma'}^{-}(f_{{\bf k}'\sigma'}-f_{{\bf k}\sigma}).
\end{align*}
Here, we assume the form of asymmetric scattering rate to be
\begin{align*}
W_{{\bf k}\sigma,{\bf k}'\sigma'}^{-}=
\left\{\begin{array}{ll}
0 & \qquad\text{(if $\sigma=\sigma'$)}\\
2\pi \sigma c (k_z-k'_z)\delta(\varepsilon_{\bf k\sigma}-\varepsilon_{\bf k'\sigma'}) & \qquad\text{(if $\sigma\ne\sigma'$)}
\end{array}\right.
\end{align*}
where $c=\frac{J^2}{N}\chi_v$ is a real constant and $\chi_v\equiv\langle({\bf S}_i\times {\bf S}_j)_z\rangle$ is the thermal average of the $z$ component of the vector spin chirality between the nearest-neighbor spins along the $z$ axis.

This asymmetric scattering term corresponds to the $k\ll1$ case of the two-spin impurity cluster in Eq.~\eqref{eq:result:Wkk}. It also applies to the paramagnetic phase of the Kondo lattice models where the correlation length between the localized moments are short. In this case, the magnetic moments in Eq.~\eqref{eq:result:Wkk} should be replaced by the thermal average,
\begin{align*}
W^-_{\vec k\sigma,\vec k'\bar\sigma}
=&\frac{2\pi J^2\sigma}{N^2}\sum_{i,j}(\vec k-\vec k')\cdot\vec r_{ij}\langle({\bf S}_i\times {\bf S}_j)_z\rangle\delta(\varepsilon_{\bf k\sigma}-\varepsilon_{\bf k'\sigma'}).
\end{align*} 
Here, the sum is over all localized moments in the system. This sum is reduced to the sum over nearest-neighbor bonds when the correlation length is similar or less than the lattice spacing, i.e., $\langle({\bf S}_i\times {\bf S}_j)_z\rangle\sim0$ for further-neighbor bonds. Assuming $\langle({\bf S}_i\times {\bf S}_j)_z\rangle=\chi_z\ne0$ only for the nearest-neighbor bonds along the $z$ axis, the constant $c$ reads $c=J^2\chi_z$ (we chose the unit of length as $|{\bf r}_{ij}|=1$ for the nearest-neighbor bonds).

We solve the Boltzmann equation in Eq.~\eqref{eq:result:boltzmann} with the scattering rate in Eq.~\eqref{eq:result:Wasym} by expanding $f_{{\bf k}\sigma}$ up to the second order in $\bf E$ and linear order in $c$~\cite{Isobe2018}; $f_{{\bf k}\sigma}=f^0_{{\bf k}\sigma}+\sum_{i=1,2,j=0,1}g^{(i,j)}_{{\bf k}\sigma}$ where $f^0_{{\bf k}\sigma}=1/(1+e^{\beta \varepsilon_{\bf k\sigma}})$ is the Fermi distribution function and $g^{(i,j)}_{{\bf k}\sigma}$ is the deviation from the equilibrium distribution in the $i$th-order in $E$ and $j$th order in $c$. We find
\begin{subequations}
\begin{align*}
g^{(1,0)}_{{\bf k}\sigma}=&-\tau e {\bf E}\cdot{\bf \nabla}_kf_{{\bf k}\sigma}^0=\tau e{\bf E}\cdot{\bf v}_{{\bf k}\sigma}\delta(\varepsilon_{{\bf k}\sigma}),\\
g^{(1,1)}_{{\bf k}\sigma}=&\tau\int\frac{d{\bf k}'}{(2\pi)^3}W^-_{{\bf k}\sigma,{\bf k}'\sigma'}(g^{(1,0)}_{{\bf k}'\sigma'}-g^{(1,0)}_{{\bf k}\sigma}),\\
g^{(2,0)}_{{\bf k}\sigma}=&-\tau e {\bf E}\cdot{\bf \nabla}_k g^{(1,0)}_{{\bf k}\sigma},\\
g^{(2,1)}_{{\bf k}\sigma}=&-\tau e {\bf E}\cdot{\bf \nabla}_k g^{(1,1)}_{{\bf k}\sigma}+\tau \int\frac{d{\bf k}'}{(2\pi)^3}W^-_{{\bf k}\sigma,{\bf k}'\sigma'}(g^{(2,0)}_{{\bf k}'\sigma'}-g^{(2,0)}_{{\bf k}\sigma}).
\end{align*}
\end{subequations}

In the Boltzmann theory, the current along the $z$ axis reads
\begin{align*}
J_z=e\sum_\sigma\int\frac{d{\bf k}}{(2\pi)^3} v_{{\bf k}\sigma}^zf_{{\bf k}\sigma}=e\sum_\sigma\sum_{i,j}\int\frac{d{\bf k}}{(2\pi)^3} v_{{\bf k}\sigma}^zg^{(i,j)}_{{\bf k}\sigma}.
\end{align*}
Here, $\rho_{\sigma}\equiv \int\frac{d{\bf k}}{(2\pi)^3}\delta(\varepsilon_{{\bf k}\sigma})$ is the density of states for the electrons with spin $\sigma$. Therefore, the nonreciprocal current in ${\cal O}(E^2)$ reads
\begin{align*}
J_z^{(2)}=e\sum_\sigma\int\frac{d{\bf k}}{(2\pi)^3} v_{{\bf k}\sigma}^z\left[g^{(2,1)}_{{\bf k}\sigma}+g^{(2,2)}_{{\bf k}\sigma}\right].
\end{align*}
The $g^{(2,1)}_{{\bf k}\sigma}$ term contributes to the nonreciprocal current when the electronic band is asymmetric due to the absence of both time and spatial inversion symmetries~\cite{Ideue2017}; this term vanishes in our case. Therefore, we here focus on the second term related to $g^{(2,2)}_{{\bf k}\sigma}$. The nonreciprocal current reads
\begin{align*}
J_z^{(2)}=&-\frac{16\pi}{5}m\tau e\rho_+\rho_- c M\frac{4\mu^2-3M^2}{\mu^2-M^2} \left(\frac{\tau e E}{m}\right)^2,\\
\sim&-\frac{144\pi}5\frac{\tau m}{e\mu^2}cM\sigma_0^2E^2.
\end{align*}
Here, $\rho_\sigma=\frac{m}{2\pi^2}\sqrt{2m(\mu+\sigma M)}$ is the density of states for the electrons with spin $\sigma$. In the second line, we assumed $\mu\gg M$ and expanded up to the leading order in $M$; $2\sigma_0=\frac{4e^2\tau\mu}{3m}$ is the linear conductivity of electrons at $M=0$ and $c=0$. Hence, the nonreciprocal current is proportional to the vector spin chirality $c=J^2\chi_z$ and magnetic polarization of the itinerant electrons $M$.

Similarly, the spin current reads
\begin{align*}
J_z^{(2)}=&\frac\hbar2\sum_\sigma\int\frac{d{\bf k}}{(2\pi)^3} \sigma v_{{\bf k}\sigma}^z\left[g^{(2,1)}_{{\bf k}\sigma}+g^{(2,2)}_{{\bf k}\sigma}\right],\\
=&-\frac{4\pi\hbar\tau^3e^2}m c\rho_+\rho_-\mu\left\{1+\frac15\left(\frac{\mu^2+3M^2}{\mu^2-M^2}\right)\right\},\\
\sim&-\frac{54\pi\hbar}5\frac{\tau m}{e^2\mu} c\, \sigma_0^2 E^2.
\end{align*}
The last equation is the result for $\mu\gg M$. The last equation implies the chiral fluctuation produces spin current in a paramagnetic phase without magnetization. A study on electric polarization by spin canting finds the polarization is parallel to ${\bf r}_{ij}\times {\bf j}_s$ where ${\bf j}_s\propto {\bf S}_i\times{\bf S}_j$ is the ``supercurrent'' of spin current~\cite{Katsura2005}. In contrast, our result finds the component of ${\bf j}_s$ parallel to ${\bf r}_{ij}$ is proportional to the spin current of electrons [Fig.~\ref{fig:intro:schematic}(a)].

\subsection*{Magnetic phase diagram}

Onsager's reaction field theory is used to calculate the magnetization and the vector spin chirality under external magnetic field~\cite{Onsager1936,Brout1967,Matsuura2003}. This method incorporates the $\sum_i|{\bf S}_i|^2=N_s$ constraint by introducing a Lagrange's multiplier $\lambda$. The effective Hamiltonian reads
\begin{align*}
H_{\rm eff}=\tilde H_{\rm cm}+\lambda\sum_i|{\bf S}_i|^2.
\end{align*}
Using this method, we find the magnetization and vector spin chirality are given by
\begin{align*}
m_z=-\frac{h}{2\lambda},
\end{align*}
and
\begin{align*}
\chi_z=\frac{2DT}{3\pi^2}\int_0^\Lambda\frac{dq\,q^4}{(Jq^2-\lambda)^2-D^2q^2}.
\end{align*}
Here, ${\bf q}=(q_x,q_y,q_z)$ is the wavenumber of the classical spin wave modes, $q=|{\bf q}|$, $\lambda$ is determined by
\begin{align*}
1-\frac{h^2}{4\lambda^2}=\int\frac{d{\bf q}}{(2\pi)^3}T\,{\rm Tr}\left(\frac1{\lambda-J_{\bf q}}\right),
\end{align*}
and
\begin{align*}
J_{\bf q}=\left(\begin{array}{ccc}
Jq^2 & {\rm i}Dq_z & -{\rm i}Dq_y \\
-{\rm i}Dq_z & Jq^2 & {\rm i}Dq_x \\
{\rm i}Dq_y & -{\rm i}Dq_x & Jq^2
\end{array}\right).
\end{align*}

This model does not show a phase transition for arbitrary choices of $h$ and $T$ when $D\ne0$. This is an artifact of the approximation used in $J_{\bf q}$, where the model has a $SO(3)$ rotational symmetry in the momentum space. In the lattice model, however, the small anisotropy due to discrete rotational symmetry breaks the $SO(3)$ symmetry. To give an idea on the ordering by the anisotropy, we defined the system is magnetically ``ordered'' if $\lambda(T,h)+D^2/(4J)<-10^{-4}$. Here, $-D^2/(4J)$ is the ground state energy. In Fig.~\ref{fig:result:MnSi}(c), we plot the ``phase boundary'' by the red solid line.

\acknowledgements
We thank T. Arima, Y. Fujishiro, N. Kanazawa, T. Morimoto, and Y. Tokura for fruitful discussions. This work was supported by JSPS KAKENHI Grant Numbers JP18H04222, JP18H03676, and JP19K14649, and JST CREST Grant Numbers JPMJCR16F1 and JPMJCR1874.

\vspace{.4cm}


\begin{thebibliography}{99}
\bibitem{Ye1999}           Ye, J., Kim, Y. B., Millis, A. J., Shraiman, B. I., Majumdar, P., \& Te\v sanovi\'c, Z. Berry phase theory of the anomalous Hall effect: Application to colossal magnetoresistance manganites. Phys. Rev. Lett. {\bf83}, 3737-3740 (1999).
\bibitem{Ohgushi2000}      Ohgushi, K., Murakami, S., \& Nagaosa, N. Spin anisotropy and quantum Hall effect in the kagom\'e lattice: Chiral spin state based on a ferromagnet. Phys. Rev. B {\bf62}, R6065-R6068 (2000).
\bibitem{Katsura2005}      Katsura, H., Nagaosa, N., \& Balatsky, A. V. Spin current and magnetoelectric effect in noncollinear magnets. Phys. Rev. Lett. {\bf95}, 057205 (2005).
\bibitem{Rikken2001}       Rikken, G. L. J. A., Folling, J., \& Wyder, P. Electrical magnetochiral anisotropy. Phys. Rev. Lett. {\bf87}, 236602 (2001).
\bibitem{Yokouchi2017}     Yokouchi, T., Kanazawa, N., Kikkawa, A., Morikawa, D., Shibata, K., Arima, T., Taguchi, Y., Kagawa, F., \& Tokura, Y. Electrical magnetochiral effect induced by chiral spin fluctuations. Nat. Commun. {\bf8}, 866 (2017).
\bibitem{Tatara2002}       Tatara, G. \& Kawamura, H. Chirality-driven anomalous Hall effect in weak coupling regime. J. Phys. Soc. Jpn. {\bf71}, 2613-2616 (2002).
\bibitem{Ishizuka2018}     Ishizuka, H. \& Nagaosa, N. Spin chirality induced skew scattering and anomalous Hall effect in chiral magnets. Sci. Adv. {\bf4}, eaap9962 (2018).
\bibitem{Nagaosa2010}      Nagaosa, N., Sinova, J., Onoda, S., MacDonald, A. H., \& Ong, N. P. Anomalous Hall effect. Rev. Mod. Phys. {\bf82}, 1539-1592 (2010).
\bibitem{Rikken2005}      Rikken, G. L.J. A., \& Wyder, P. Magnetoelectric anisotropy in diffusive transport. Phys. Rev. Lett. 94, 016601 (2005).
\bibitem{Tokura2018}      Tokura, Y., \& Nagaosa, N. Nonreciprocal responses from non-centrosymmetric quantum materials. Nat. Commun. {\bf9}, 3740 (2018).
\bibitem{Onsager1931}      Onsager, L. Reciprocal relations in irreversible process. I. Phys. Rev. {\bf37}, 405-426 (1931).
\bibitem{Kubo1957}         Kubo, R. Statistical-mechanical theory of irreversible process. I. General theory and simple applications to magnetic and conduction problems. J. Phys. Soc. Jpn. {\b12}, 570-586 (1957).
\bibitem{Fisher1981}       Fisher, D. S., \& Lee, P. A. Relation between conductivity and transmission matrix. Phys. Rev. B {\bf23}, 6851-6854 (1981).
\bibitem{Pop2014}          Pop, F., Auban-Senzier, P., Canadell, E., Rikken, G. L. J. A., \& Avarvari, N. Electrical magnetochiral anisotropy in a bulk chiral molecular conductor. Nat. Commun. {\bf5}, 3757 (2014).
\bibitem{Ideue2017}        Ideue, T., Hamamoto, K., Koshikawa, S., Ezawa, M., Shimizu, S., Kaneko, Y., Tokura, Y., Nagaosa, N., \& Iwasa, Y. Bulk rectification effect in a polar semiconductor. Nat. Phys. {\bf13}, 578-583 (2017).
\bibitem{Wakatsuki2017}    Wakatsuki, R., Saito, Y., Hoshino, S., Itahashi, Y. M., Ideue, T., Ezawa, M., Iwasa, Y., \& Nagaosa, N. Nonreciprocal charge transport in noncentrosymmetric superconductor. Sci. Adv. {\bf3}, e1602390 (2017).
\bibitem{Avci2015}         Avci, A. O., Garello, K., Ghosh, A., Gabureac, M., Alvarado, S. F., \& Gambardella, P. Unidirectional spin Hall magnetoresistance in ferromagnet/normal metal bilayers. Nat. Phys. {\bf11}, 570-575 (2015).
\bibitem{Yasuda2016}       Yasuda, K., Tsukazaki, A., Yoshimi, R., Takahashi, K. S., Kawasaki, M., \& Tokura Y. Large unidirectional magnetoresistance in a magnetic topological insulator. Phys. Rev. Lett. {\bf117}, 159903 (2019).
\bibitem{Aoki2019}         Aoki, R., Kousaka, Y., \& Togawa, Y. Anomalous Nonreciprocal Electrical Transport on Chiral Magnetic Order, Phys. Rev. Lett. {\bf122}, 057206 (2019).
\bibitem{Ishikawa1976}     Ishikawa, Y., Tajima, K., Bloch, D., \& Roth, M., Helical spin structure in manganese silicide MnSi. Sol. Stat. Commun. {\bf19}, 525-528 (1976).
\bibitem{Bak1980}          Bak, P., \& Jensen, M. H. Theory of helival magnetic structures and phase transition in MnSi and FeGe. J. Phys. C: Sol. Stat. Phys. {\bf13}, L881-885 (1980).
\bibitem{Roessler2006}     R\"o{\ss}ler, U. K., Bogdanov, A. N., \& Pfleiderer, C. Spontaneous skyrmion ground states in magnetic metals. Nature {\bf442}, 797-801 (2006).
\bibitem{Muhlbauer2009}    M\"uhlbauer, S., Binz, B., Jonietz, F., Pfleiderer, C., Rosch, A., Nuebauer, A., Goergii, R., \& B\"oni, P. Skyrmion lattice in a chiral magnet. Science {\bf323}, 915-919 (2009).
\bibitem{Yu2010}           Yu, X. Z., Onose, Y., Kanazawa, N., Park, J. H., Han, J. H., Matsui, Y., Nagaosa, N., \& Tokura, Y. Real-space observation of a two-dimensional skyrmion crystal. Nature {\bf465}, 901-904 (2010).
\bibitem{Isobe2018}        Isobe, H. \& Fu, L. High-frequency rectification via chiral Bloch electrons. preprint (arXiv:1812.08162) (2018).
\bibitem{Onsager1936}      Onsager, L. Electric moments of molecules in liquids. J. Am. Chem. Soc. {\bf58}, 1486-1493 (1936).
\bibitem{Brout1967}        Brout, R., \& Thomas, H. Molecular field theory, the Onsager reaction field and the spherical model. Physics {\bf3}, 317-329 (1967).
\bibitem{Matsuura2003}     Matsuura, M., Endoh, Y., Hiraka, H., Yamada, K., Mishchenko, A. S., Nagaosa, N. \& Solovyev, I. V. Classical and quantum spin dynamics in the fcc antiferromagnet NiS$_2$ with frustration. Phys. Rev. B {\bf68}, 094409 (2003).
\bibitem{Jeong2004}        Jeong, T. \& Pickett, W. E. Implications of the B20 crystal structure for the magnetoelectronic structure of MnSi. Phys. Rev. B {\bf70}, 075114 (2004).
\bibitem{Lee2007}          Lee, M., Onose, Y., Tokura, Y. \& Ong, N. P. Hidden constant in the anomalous Hall effect of high-purity magnet MnSi. Phys. Rev. B {\bf75}, 172403 (2007). 
\bibitem{Ishizuka2017}     Ishizuka, H. \& Nagaosa, N. Noncommutative quantum mechanics and skew scattering in ferromganetic metals. Phys. Rev. B {\bf96}, 165202 (2017).
\bibitem{Ishizuka2018b}    Ishizuka, H. \& Nagaosa, N. Impurity-induced spin chirality and skew scattering in magnetic metals. New J. Phys. {\bf20}, 123027 (2018).
\end{thebibliography}
\end{document}